# Slow dynamic elastic recovery in unconsolidated metal structures

John Y. Yoritomo and Richard L. Weaver
Department of Physics
University of Illinois, Urbana, IL 61801

**Abstract**

Slow dynamic nonlinearity is widely observed in brittle materials with complex heterogeneous or cracked microstructures. It is seen in rocks, concrete and cracked glass blocks. Unconsolidated structures show the behavior as well: aggregates of glass beads under pressure and a single glass bead confined between two glass plates. A defining feature is the loss of stiffness after a mechanical conditioning, followed by a logarithmic-in-time recovery. Materials observed to exhibit slow dynamics are sufficiently different in microstructure, chemical composition, and scale (ranging from the laboratory to the seismological) to suggest some kind of universality. There lacks a full theoretical understanding of the universality in general and the log(time) recovery in particular; one suspicion has been that the phenomenon is associated with glassy grain boundaries and microcracking. Seminal studies were focused on sandstones and other natural rocks, but in recent years other experimental venues have been introduced with which to inform theory. Here, we present measurements on some simple *metallic* systems: an unconsolidated aggregate of aluminum beads under a confining pressure, and an aluminum bead confined between two aluminum plates, and a steel bead confined between steel plates. Ultrasonic waves are used as probes of the systems, and changes are assessed with coda wave interferometry. Three different methods of low-frequency conditioning are applied; all reveal slow dynamic nonlinearities. Results imply that glassy microstructures and cracking do not play essential roles, as they would appear to be absent in our systems.

## I. Introduction

The past two decades have seen numerous reports of what has come to be termed slow dynamic nonlinear elasticity in which materials exhibit a loss of stiffness after a mechanical or thermal conditioning, followed by a slow, log(time), recovery [1–6]. This behavior is readily observed in natural rocks, in cements, in concretes and in cracked glass, all of which possess an internal geometry composed of many contact points with glassy microstructure. (One exception may be a report by Johnson and Sutin [4] of slow dynamics in a pearlite/graphite (grey iron) composite and a sintered steel that presumably had complex microstructures of their own.) In spite of the ubiquity of these behaviors, they are poorly understood. It may be that glassy microstructures are essential. This notion is motivated in part by the well-known slow dynamics of electronic properties in metallic glasses, materials with a complex energy landscape that can provide a broad spectrum of barrier energies that in turn leads to 1/f noise and log(t) like evolutions [7]. Darling *et al.* [8] showed, using neutron diffraction in a sandstone, that the crystallites were behaving linearly. They concluded that the peculiar behavior was confined to the glassy grain contacts.

Slow dynamic nonlinearity is also found in unconsolidated materials. It has been observed in aggregates of spherical glass beads, by Johnson and Jia [9], by Zaitsev *et al.* [10], and by Yoritomo and Weaver [11]. The latter also observed slow dynamics in a single glass bead confined between two glass plates [12]. These materials are clearly glassy. They may also be microcracked, as the tensile stresses near the contact points exceeded nominal strengths of glass. So it is unclear if the slow dynamics in these systems was due to processes on internal crack surfaces (as it surely is in tests on cracked glass blocks) or to processes at the gross glass/glass interfaces. Furthermore, as the tests were in glass, they did not challenge the idea that glassy microstructures are critical.

Here we repeat the measurements of Yoritomo and Weaver [11,12] but replace their glass beads with aluminum and steel. The presumption is that while these materials may yield near their contacts, they do not generate cracks. The other presumption is that there are no glassy microstructures. Thus any observed behaviors must be explained without reference to cracks or glass.

The following three sections respectively describe our tests on an aluminum bead pack under a confining static pressure, our tests on a single aluminum bead between two aluminum plates, and our tests on a single hardened steel ball bearing between two steel plates.

**II. Aluminum Bead Pack**

The structure is identical to that described earlier [11], except that instead of 3mm diameter soda-lime glass beads, we use monodisperse $2r = 3.1$mm diameter aluminum beads of unknown alloy. The aggregate of 229g is composed of about 5400 such beads, and fills a cylindrical volume of 33mm thickness, and 71mm diameter. It is confined vertically between two thin steel disks, and laterally by an annulus of high strength polystyrene foam. A dead-weight pressure of 109kPa is created by 44kg of steel slabs. Vertical forces are transmitted from the slabs to the bead pack, and from the bead pack to the laboratory floor through thick-walled hollow cylinders of high strength polystyrene foam. Ultrasonic transducers are placed on the top and bottom steel disks.

Hertzian theory (see Appendix I) predicts a shear stress maximum of 229MPa at a point about 10μ from the center of bead-bead contacts. This is comparable to nominal yield stress (155-243MPa) in heat-treated aluminum. We therefore expect some plastic flow, there, and at asperities. Plastic flow at crack tips will prevent any incipient microcracks from propagating, so we do not expect cracking.

As previously, the transmitted signal has a high-frequency cutoff, now near 150kHz, that we associate with the upper band edge of rotational waves in the aggregate. The received signal has a time-domain envelope characteristic of wave energy diffusion. See Appendix I for discussion.

The bead pack is mechanically conditioned (or "pumped") in three different ways: impulsively with an impact, harmonically with a many-cycle sinusoidal force, and quasi-statically.

Figure 1 shows the observed signal "stretches" obtained by the coda-wave interferometry methods discussed in refs. [11,13,14] after a transient pump force provided by dropping a rubber ball on top of the structure. The negative values correspond to signal delays and loss of stiffness relative to times before impact. We estimate (see Appendix I) peak strain in the bead pack to be $\epsilon_{peak} =$ $28 \times 10^{-6}$ occurring about 10ms after the impact. The slope $d\text{"stretch"}/d\ln(t) = m = 19 \times 10^{-5}$ of the recovery (see inset of Fig. 1) divided by the peak strain is 6.8, greater than that seen in the 215kPa glass bead pack. See Table 1 for a summary. The accelerated slope after $\sim e^5$ seconds is plausibly attributable to remnant recoveries from earlier pumpings.

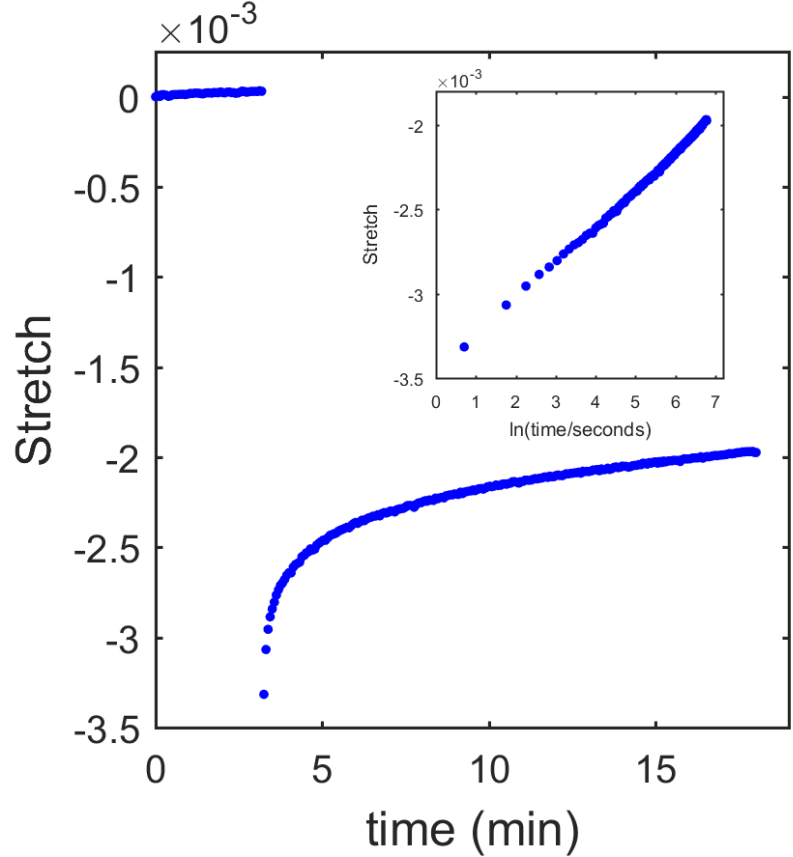

Figure 1. After the ball drop, which generated a peak strain in the bead pack of $28 \times 10^{-6}$, the received signal is slower than before, by about 0.33%. It then recovers like log(t), as shown in the inset.

Figure 2a shows observed signal stretches during and after harmonic conditioning. Three on/off cycles in which an electro-magnetic vibration shaker placed on the top of the structure was driven at 60Hz for two minutes, followed by two minutes of no shaking. The resulting harmonic strain amplitude is estimated to be $\epsilon_{rms} = 0.65 \times 10^{-6}$ (see Appendix I). As seen in Fig. 3 (blue dots) the recovery slope is $m = 0.27 \times 10^{-4}$, leading to $m/\epsilon_{rms} = 42$.

Figure 2b shows the observed signal stretches while an extra $\Delta M = 1$kg mass is quasi-statically (QS) added and then removed, three successive times, from the top of the structure. The resulting strain in the bead pack is estimated to be $\epsilon = 2.35 \times 10^{-6}$ (see Appendix I). As before, we remark that the recovery process after a mass is removed or added is such that the structure grows stiffer with time, regardless of the sign of the pumping. Recovery slopes $m$ are $3.5 \times 10^{-5}$ after adding the weight, and slightly less after removing the weight. After normalizing, this is $m/\epsilon = 15$.

Table 1 summarizes and compares slow dynamic recoveries in the glass and aluminum bead packs for the three kinds of pumping.

We have also done tests on the aluminum bead pack at 215kPa (see Appendix II), and found the slow dynamic recoveries to be of similar magnitude but far less clean than those shown in Figs. 1-3 or those of the glass bead pack [11] at 215kPa. At this load, the Al bead pack exhibits many slips minutes or more after the pump, and these contaminate the stretch versus time plots. It is also plausible to attribute the difference to extra plastic flow at the contact points.

In sum, the aluminum bead pack at 109 kPa behaves very much as did the glass bead pack, with similar recovery slopes for similar pump strains. Quantitative differences are not large and could plausibly be ascribed to the different static load. It is remarkable that, in spite of the very different material properties (other than elastic modulus and density and bead size) of aluminum and glass, the slow dynamic behaviors have rather similar amplitudes.

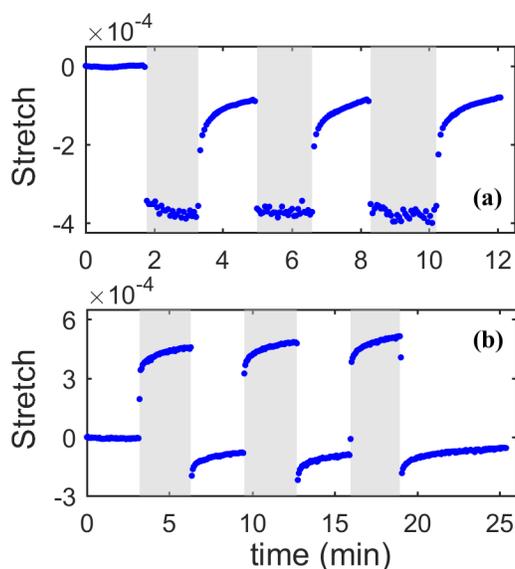

Figure 2. Panel (a) shows three cycles of harmonic conditioning (shaded regions) with rms strain $0.65 \times 10^{-6}$ on the 109 kPa aluminum bead pack. Panel (b) shows quasi-static conditioning, where three cycles of adding (shaded region) and removing an extra 1 kg mass to the top of the structure correspond to strain changes of $2.35 \times 10^{-6}$. Both show the characteristic recoveries.

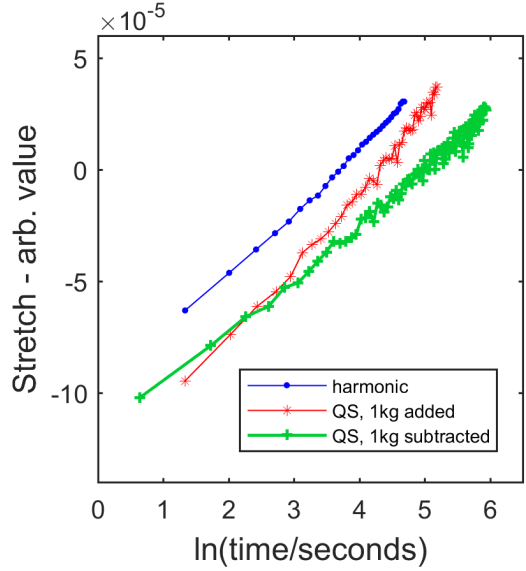

Figure 3. The last period of harmonic recovery (blue dots), the last period of quasi-static conditioning (red asterisks), and the last period of quasi-static recovery (green crosses) plotted vs. ln(time) for the aluminum bead pack. As seen in the glass bead pack, the slope is greater on loading than unloading in the aluminum one, though the difference is diminished. For harmonic recovery and quasi-static conditioning, the slight uptick after $e^4$ seconds is plausibly attributed to temperature drifts and/or contributions from recoveries associated with earlier pumpings.

| Structure & pump style | Pump strain $\times 10^{-6}$ | Slope $m$ $\times 10^{-5}$ | $m/\epsilon$ |
|---|---|---|---|
| Glass bead pack, QS pump | $\Delta\epsilon = 2.1$ | 2.4 | 11.5 |
| Glass bead pack, Impulse pump | $\epsilon_{peak} = 28$ | 4.8 | 1.7 |
| Glass bead pack, Harmonic pump | $\epsilon_{rms} = 0.62$ | 1.4 | 23 |
| | | | |
| Al bead pack, QS pump | $\Delta\epsilon = 2.35$ | 3.5 | 14.9 |
| Al bead pack, Impulse pump | $\epsilon_{peak} = 28$ | 19 | 6.8 |
| Al bead pack, Harmonic pump | $\epsilon_{rms} = 0.65$ | 2.7 | 42 |

Table 1. Comparison of slow dynamic recoveries in the 215kPa glass bead pack and the 109kPa aluminum bead pack. The slope estimates for quasi-static (QS) pumping are for adding weight.

### III. Single Bead Systems: Aluminum and Steel

We also study slow dynamics in a single bead systems, with structures similar to the single glass bead system described earlier [12]. The first system is a single $2r = 3.1$mm diameter 2017T4 aluminum bead confined between two aluminum alloy (6061 T651) plates of mass 1.71kg and dimensions 215mm x 155mm x 19mm. The second system has one $2r = 3$mm diameter hardened steel bead confined between two steel plates, of mass 4.043kg (101mm x 100mm x 49mm) on the bottom, and 5.107kg (253mm x 105mm x 26mm) on the top.

The weight of the upper plate and the geometry of the structure implies, using simple Statics, a normal contact force across both the aluminum and steel beads of $F = 3.4$N. (Larger forces lead to plots of delay versus signal time contaminated by numerous features that we attribute to aftershocks.) Hertzian theory (see Appendix I) estimates a peak shear stress of 239 MPa in the aluminum and 426MPa in the steel. Both are comparable to nominal yield shear stress in heat treated aluminum alloys (155-243MPa) and steel (200-850MPa). Thus there may be some plastic flow, but we do not expect cracking.

As previously [12], we observe a diffuse signal in the lower plate that rises slowly before dissipating. This transmitted wave is strongest in a narrow pass band, near 120kHz for the aluminum system and 105kHz for the steel. We associate that resonant transmission with the two-fold degenerate rigid-body rotational modes of the bead. The aluminum (steel) waveform is further filtered into the band from 80 to 140kHz (130kHz) before further processing. See Appendix I for further discussion.

The confined aluminum bead is mechanically pumped in three different ways: impulsively, harmonically with a many-cycle sinusoidal force, and quasi-statically (QS). The steel bead is conditioned only impulsively and quasi-statically.

Figure 4 shows the observed lower-slab signal delay, or average shift (blue dots in Fig. 4 for aluminum and red crosses for steel),, obtained by coda-wave interferometry [13,14] before and after an impulsive pump provided by dropping a 7g rubber ball from a height of 10cm onto the top of the structure. As discussed previously [12], the change in transmitted wave is not well described by a simple stretch like it was in the bead pack, but rather as an apparent stretch for very early times followed by an irregular fixed delay τ. The delays τ in each of about fifty nonoverlapping 200µ$s$ windows from 0.05 to 10ms are averaged and the result is plotted versus laboratory time, before and after the impact. The inset of Fig. 4 shows that the changes recover logarithmically. The peak of the strain across the bead is estimated to be $\epsilon_{peak} = 1360 \times 10^{-6}$ for aluminum and $470 \times 10^{-6}$ for steel (see Appendix I). These are enormous strains by the standards of the slow dynamics literature. Scrutiny of the accelerometer signal after the impacts shows the vibrations to be in a nonlinear regime that renders problematic any normalization by peak pump strain. Nevertheless we display the resulting recoveries in fig 4, because they have a good signal to noise ratio and the aluminum recovery exhibits an interesting blip at ln(t) = 3.7

Within the inset of Fig. 4 we expand the recovery of the aluminum bead near ln(t) = 3.7. It is similar to several events seen in the 215 kPa aluminum bead pack after pumping. We hypothesize that it is an aftershock, with its own loss of stiffness and its own slow dynamic recovery. We did not observe such events in the glass bead measurements.

In order to report measurements in the regime of linear vibrations we repeated the test with a smaller, 720 mg, wooden ball. For this case the peak strain for the aluminum system was $170 \times 10^{-6}$ and the recovery slope was 0.0166 μsec, for a normalized recovery slope of 97 μs. For the steel system, peak strain was $190 \times 10^{-6}$, recovery slope 0.012μs, and normalized slope 63μs. The recoveries after this smaller impulse for both systems are plotted in Appendix III.

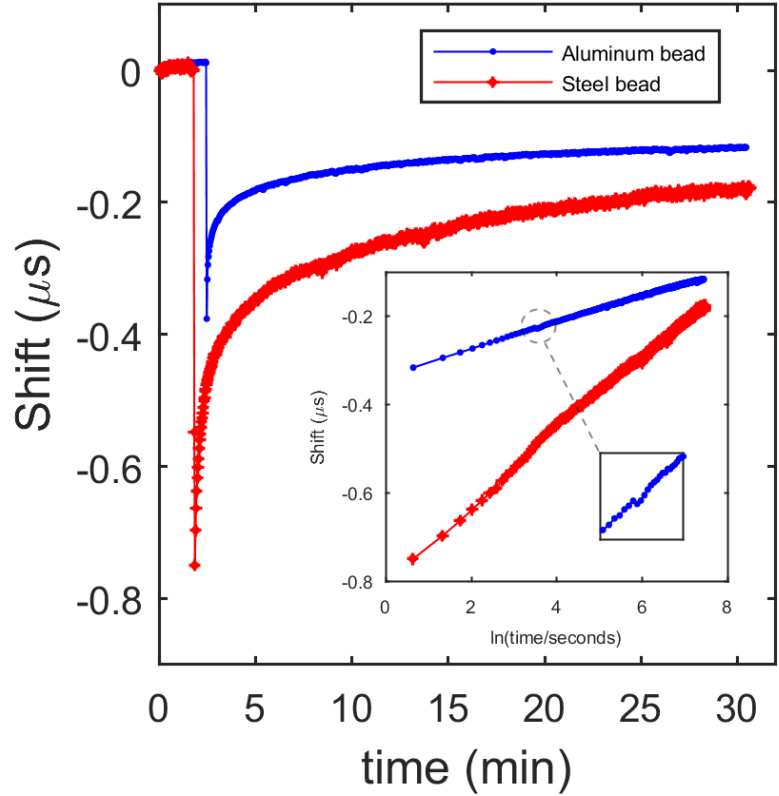

Figure 4. Slow dynamic recovery of the shift (negative of delay) in the transmitted signal through a single aluminum bead (blue dots) and through a single steel bead (red crosses) after pumping by a dropped 7g ball. Both recoveries are logarithmic-in-time (inset). The odd event at ln(t) = 3.7 in the aluminum bead recovery (expanded in second inset) is similar to several events seen in the 215 kPa aluminum bead pack after pumping.

Figure 5a shows the mean shifts for the aluminum system before and after 200Hz harmonic conditioning from placing an electro-magnetic shaker on the bottom slab. We estimate the harmonic strain amplitude across the bead to be $\epsilon_{rms} = 44 \times 10^{-6}$ (see Appendix I). Again we see log(time) recovery (Fig. 6., green crosses). There is a noisiness here (even when the shaker is off) that is not present in the earlier report . This is due to our lack of access to the optical vibration isolation table used previously. The slow dynamic recovery is evident regardless.

Figure 5b shows the mean shifts (blue dots connected by solid line for the aluminum and red dots connected with dash-dotted line for steel) before and after alternately placing and removing a 65g weight on the upper plate over the bead. Based on Hertzian theoryfor the stiffness of the contacts (see Appendix I), the corresponding quasi-static strain changes across the bead are $110 \times 10^{-6}$ for aluminum and $60 \times 10^{-6}$ for steel. The shifts vs. log(time) for both adding and removing the 65g weight are plotted in Fig. 6 (red asterisks and blue dots for aluminum, and cyan asterisks and black dots for steel).

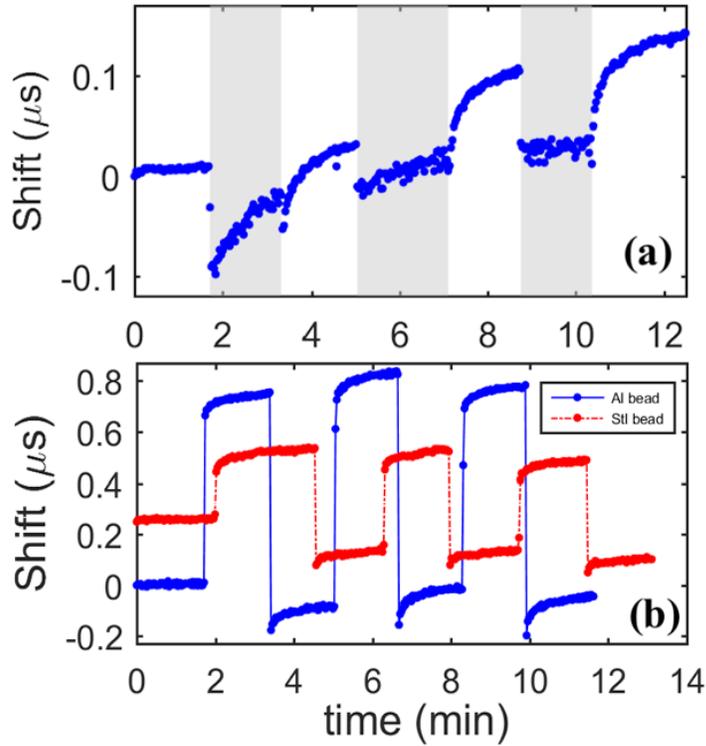

Figure 5. Panel (a) shows the response of the single aluminum bead system to harmonic pumping with rms strain $44 \times 10^{-6}$. Panel (b) shows the response of the single aluminum bead system (blue dots, connected with solid line) and the single steel bead system (red dots, connect with a dash-dotted line) for strains of $110 \times 10^{-6}$ and $60 \times 10^{-6}$, respectively, when a 65g mass is added to the top plate. An overall shift of $0.25 \mu s$ has been added to the steel bead response for better clarity in the plot. There is a noisiness here (even when the pump is off) that is not present in the earlier report. The slow dynamic recovery is evident regardless.

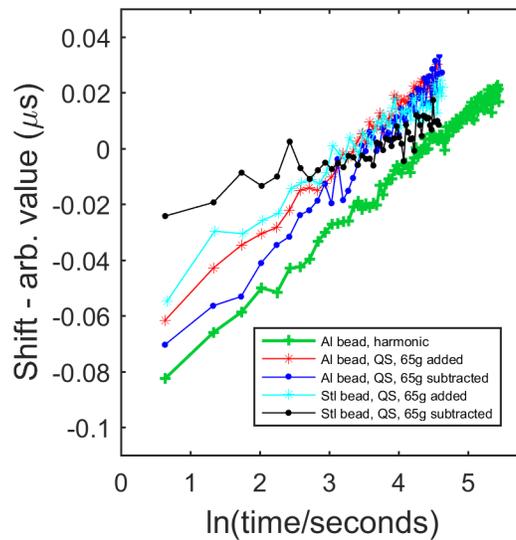

Figure 6. Recoveries of the single aluminum bead system and single steel bead system in harmonic and quasi-static pumping. The recovery of the aluminum bead to harmonic conditioning corresponds to the last recovery in Fig. 5a. The relaxation shown, due to adding and subtracting 65g from the load on both the aluminum and steel bead, corresponds to the last cycle in Fig. 5b.

Table 2 summarizes the results from these five single bead measurements in aluminum and steel. It also, for comparison, displays the earlier results for the single glass bead. The fourth column displays the slopes after normalizing by the pump strains. But the reader must be advised that we have not confirmed that our slopes are, as seen elsewhere, linear in pump strain. The normalized values are only a rough guide.

| Material & pump style | Pump strain $\times 10^{-6}$ | Slope $m'$ | $m'/\epsilon$ |
|---|---|---|---|
| QS on glass bead | $\Delta\epsilon = 120$ | 0.06µs | 500µs |
| Impulse on glass bead (90mg wood) | $\epsilon_{peak} = 57$ | 0.0095µs | 167µs |
| Harmonic on glass bead | $\epsilon_{rms} = 8.4$ | 0.026µs | 3150µs |
| | | | |
| QS on Al bead | $\Delta\epsilon = 110$ | 0.025µs | 227µs |
| Impulse on Al bead (720mg wood) | $\epsilon_{peak} = 170$ | 0.017µs | 97µs |
| Impulse on Al bead (7g rubber) | $\epsilon_{peak} = 1360$ | 0.03µs | 22µs |
| Harmonic on Al bead | $\epsilon_{rms} = 44$ | 0.017µs | 386µs |
| | | | |
| QS on steel bead | $\Delta\epsilon = 60$ | 0.015µs | 250µs |
| Impulse on steel bead (720mg wood) | $\epsilon_{peak} = 190$ | 0.012µs | 63 µs |
| Impulse on steel bead (7g rubber) | $\epsilon_{peak} = 470$ | 0.083µs | 175µs |

Table 2. Comparison and summary of slow dynamic response in three single bead systems. The static force on all three beads (glass, aluminum, and steel) was 3.4N. For impulsive pumping on the aluminum and steel beads, we report numbers for the 720mg wooden impactor dropped from 21 cm, and for the 7g rubber impactor from 10 cm that led to nonlinear vibrations. Impulsive pumping on the glass bead was done with a 90 mg wood ball from 21 cm. Slope $m' = d"shift"/dln(t)$.

The recoveries after quasi-static pumpings in the three materials are consistent. The strains are comparable (the difference being simply a matter of different effective contact stiffnesses, traceable in turn to the different Young's moduli.) The slopes of the slow dynamic recovery are comparable also, as are their normalized values. However, the reader is advised that exact comparisons between different materials are problematic: shifts scale with intrinsic stretch in the bead times the dwell time of a wave in the bead. The latter quantity is not known and may well be different for the different materials. Nevertheless, a consistent picture is developed; slow dynamic amplitudes from QS pumping are comparable, as they were in the bead packs.

The impulsive cases are also consistent. Peak pump strains differ somewhat between the metal beads and the glass bead. This should be unsurprising; peak strain is a complex function of the structure's vibration modes, in turn dependent on varying masses and supports. The slopes in the three (wooden ball) cases are similar. A consistent picture is seen; slow dynamic amplitudes from (small wooden ball) impulsive pumping are indeed comparable, as they were for the bead packs and for the QS pumpings.

Harmonic pumping was applied only to aluminum and glass single beads, as geometric constraints prohibited placement of the shaker for the steel bead. The harmonic strains developed in the aluminum and glass bead systems are rather different (again attributable to the very different supports affecting the structure's vibration modes). The recovery slopes are rather similar, but after normalizing they are very different. It is possible that our estimates for strain may be erroneous or that the normalization we use is problematic. Or it may be that slow dynamics due to harmonic pumping has a much greater amplitude in a single glass bead than in the single aluminum bead, but the similarities seen in the other structures and other pumpings suggests not.

In sum, we clearly see slow dynamic recoveries in single metal beads as well as glass beads. Quantitative comparisons suggest they have very similar slow dynamic amplitudes.

## IV. Discussion

These results have implications in the search for a microphysical basis for Slow Dynamics. It would seem clear that the slowness cannot be attributed to the diffusivity of heat as Zaitsev *et al*. [15] posited. They showed that heat conduction around cracks in glass could be slow enough to explain slow dynamics; they also showed that log(t) behavior could arise naturally. However, diffusivity of heat is much faster in metals than in silica glasses, so it would be difficult to maintain that it could be responsible for such similar behavior in both materials. It would also seem clear that, as slow dynamics occurs for aluminum and steel that lack microcracking at these loads, cracks are not essential. The only requirement is for the presence of interfaces such as exist between the beads or between the beads and their confining slabs.

We are struck by the very similar amplitudes ($m/\epsilon$) for the different materials. We are inclined to infer that the physical mechanisms for slow dynamics in metals and glass are the same. This is surprising, as the conventional sources of nonlinearity in these materials (microcracking and plasticity, glassy dynamics) are so different. Why then should their slow dynamic nonlinearities have such similar amplitudes?

The reported results here cannot pretend to be a thorough quantitative comparison of slow dynamics in glass and metals, but they are sufficiently striking that we are drawn to publish them as is. More thorough surveys, with control of alloy, heat treatment, tempering, static load, surface treatments, dynamic strain, humidity and/or temperature, and with better vibration isolation, are indicated. The laboratory structures used here are well suited to such investigations.

**Appendix I: Waveforms and spectra, Hertzian estimates, and estimates for pump strains**

Waveforms and spectra for the three different systems

*Unconsolidated aluminum bead pack*

Figure S1 shows a typical received signal, and its spectrum, on the lower disk after propagation through the aluminum bead pack. The signal has its source in an impulsive charge applied to the ultrasonic transducer on the upper disk. As previously [11], we observe a high-frequency cutoff, now near 150kHz, that we associate with the upper band edge of rotational waves in the aggregate. (Simple theory that assumes the structure is hexagonal close packed(hcp) predicts a cutoff at 187kHz. The difference is similar to that observed in the glass bead pack and attributable to the non-hcp character of the packing or perhaps to plasticity corrections to Hertzian theory.) Our signal energy is dominated by frequencies near 150kHz. The received signal has a time-domain envelope characteristic of wave energy diffusion.

*Single aluminum bead system*

Figure S2 shows a typical received signal and its spectrum in the lower aluminum alloy (6061 T651) plate of mass 1.71kg and dimensions 215mm x 155mm x 19mm after an ultrasonic impulse is applied to the top identical plate. The signal propagated from the upper slab through the 3.1mm 2017T4 aluminum bead to the transducer on the bottom slab. As previously [12], we observe a simple diffuse decay (not shown here) in the upper plate. In the lower plate we observe a diffuse signal (Fig. S2a) that rises slowly before dissipating. As previously, we observe this transmitted wave to be strongest in a narrow pass band, now near 120kHz (Fig. S2b). We associate that resonant transmission with the two-fold degenerate rigid-body rotational modes of the bead. Our signal in the lower plate is dominated by these waves. The waveform is further filtered into the band from 80 to 140kHz before further processing.

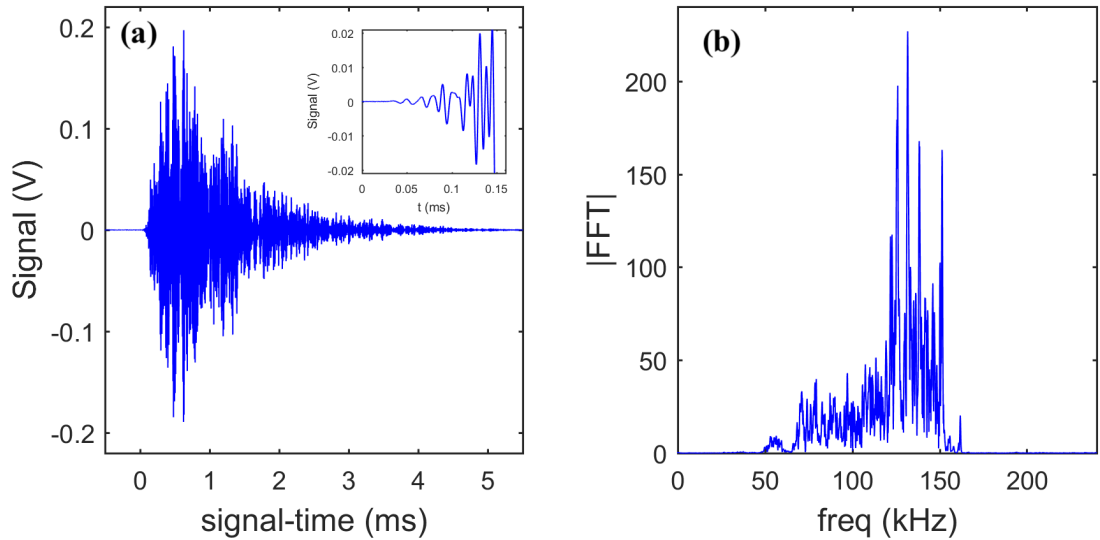

Figure S1. The signal (panel (a)) and spectrum (panel (b)) received on the bottom steel disk after an impulse is applied to the transducer on the upper disk. Signal is repetition averaged 100 times. As in the glass bead pack, the inset of panel (a) shows a first arrival at about 42.5 microseconds (somewhat slower than seen in the glass beads, a difference that could be ascribed to the lower static load). The first arrival is followed by a random looking irregular process under a diffusion envelope. As with the glass beads, we observe a sharp cutoff, here at about 150 kHz (panel (b)).

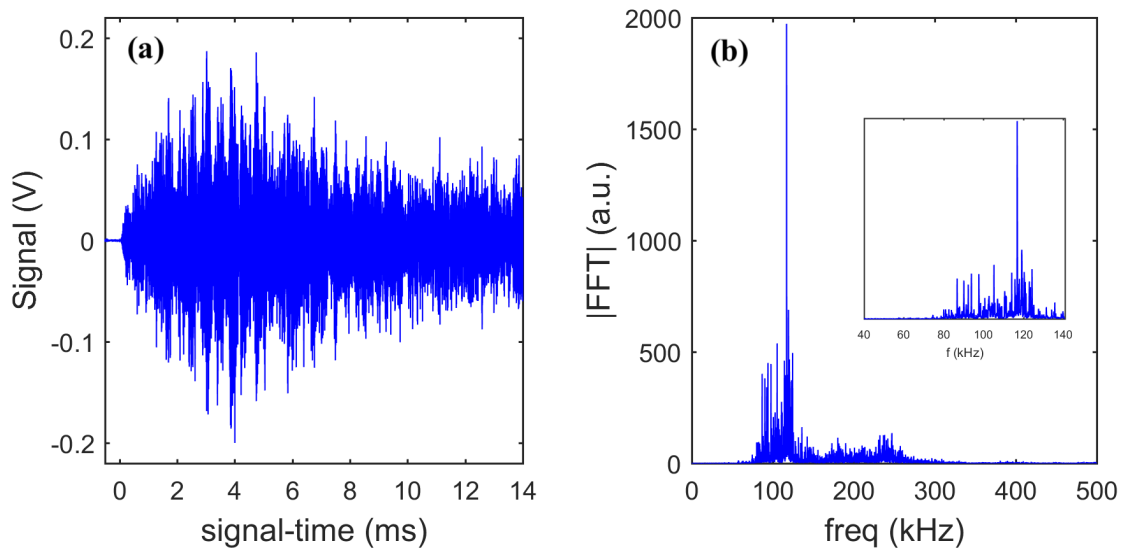

Figure S2. The signal (panel (a)) transmitted through aluminum bead into lower slab and its spectrum (panel (b)). The signal has been repetition averaged 20 times. We observe the transmitted wave to be strongest in a narrow pass band, now near 120kHz. We associate that resonant transmission with the two-fold degenerate rigid-body rotational modes of the bead. The inset of panel (b) is a close-up of the spectrum near the resonant frequency.

*Single steel bead system*

Figure S3 shows a typical received signal and its spectrum on the lower steel plate (of mass 4.043kg and dimensions 101mm x 100mm x 49mm) after an ultrasonic impulse is applied to the top plate (of mass 5.107kg and dimensions 253mm x 105mm x 26mm). The signal propagated from the upper slab through the 3mm hardened steel bead to the bottom slab. As previously, we observe a simple diffuse decay (not shown) in the upper plate. But in the lower plate we observe (Fig. S3a) a much weaker diffuse signal that rises slowly before dissipating. As previously [12], we observe (Fig. S3b) the transmitted wave to be strongest in a narrow pass band, now near 105 kHz. We associate that resonant transmission with the two-fold degenerate rigid-body rotational modes of the bead. Our signal energy in the lower plate is dominated by these waves. The signal is further band-pass filtered between 80 and 130 kHz before further processing.

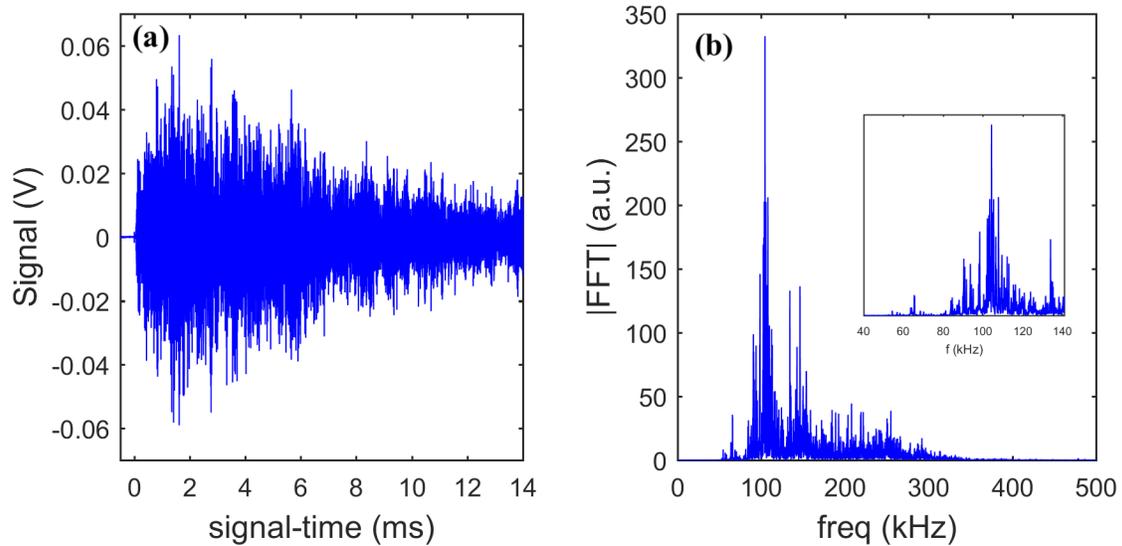

Figure S3. The signal (panel (a)) and spectrum (panel (b)) in the lower steel plate after transmission through the steel bead. The signal has been repetition averaged 20 times. We observe the transmitted wave to be strongest in a narrow pass band, now near 105kHz. We associate that resonant transmission with the two-fold degenerate rigid-body rotational modes of the bead. The inset of panel (b) is a close-up of the spectrum near the resonant frequency.

Herztian estimates for peak stresses

*Unconsolidated aluminum bead pack*

The pressure on the bead pack from the static load is $p = 109$kPa. This corresponds to normal contact forces between the beads of order $F = \pi r^2 p = 0.82$N (compared with the 215kPa glass bead pack [11] for which the contact forces were 1.5N.) Hertzian theory [16,17], using a Young's modulus of 70GPa, then predicts circular contact regions of radius $a = 23\mu$. A peak compressive stress of 755MPa (mostly hydrostatic) occurs at the center of the contacts, peak tensile stress, 83MPa, occurs at the edge of the contact circle. Shear stress is maximum (229MPa) at a point about 10μ below the center of contact. This is comparable to nominal yield stress (155-243MPa) in heat-treated aluminum. We therefore expect some plastic flow there, as well as at any rough surface asperities. We do not expect cracking.

*Single aluminum bead system*

The weight of the upper aluminum plate and the geometry of the structure implies, using simple Statics, a normal contact force across the aluminum bead of $F = 3.4\text{N}$. Hertzian theory, using a Young's modulus of 70GPa, then predicts a circular contact region of radius $a = 46\mu$, and a peak compressive stress of 771MPa, a peak tensile stress of 87.4MPa, and a peak shear stress of 239 MPa that is comparable to nominal yield shear stress (155-243MPa) in heat treated aluminum alloys. Thus there may be some plastic flow in this single bead system, but we do not expect cracking.

*Single steel bead system*

The weight of the upper steel plate and the geometry of the structure implies, again using simple Statics, a normal contact force $F$ across the steel bead of $F = 3.4\text{N}$. (Larger forces lead to plots of delay versus signal time contaminated by numerous features that we attribute to aftershocks.) Hertzian theory, using a Young modulus of 190GPa, then predicts a circular contact region of radius $a = 33\mu$, and a peak compressive stress of 1490MPa, a peak tensile stress of 209MPa, and a peak shear stress of 462MPa. This may be compared to nominal yield stress for steel in shear of 200 to 850MPa.

Estimates for pump strains

*Unconsolidated aluminum bead pack*

The bead pack is mechanically conditioned (or "pumped") in three different ways: impulsively with an impact, harmonically with a many-cycle sinusoidal force, and quasi-statically.

As before, to estimate the peak strain from impulsive conditioning, an accelerometer is placed on the top of the structure during a ball drop. The peak signal from this accelerometer, the known mass of the dead weight, and an estimated modulus for the bead pack $E = \rho c^2$ with a measured mass density of $\rho = 1750\text{kg/m}^3$ and $c = $ the apparent low frequency coherent wave speed of $33\text{mm}/42.5\mu s = 775\text{m/s}$ (slightly slower than seen in the 215kPa glass bead pack), gives us an estimated peak strain in the bead pack of $\epsilon_{peak} = 32 \times 10^{-6}$. This occurs about 10ms after the impact (compare $\epsilon_{peak} = 28 \times 10^{-6}$ seen in the glass bead pack).

Harmonic conditioning consisted of three on/off cycles in which an electro-magnetic vibration shaker placed on the top of the structure was driven at 60Hz for two minutes, followed by two minutes of no shaking. The resulting harmonic strain amplitude $\epsilon_{rms} = 0.74 \times 10^{-6}$ in the bead pack was estimated, as previously, from measurements from an accelerometer on the top of the structure, the known mass of the dead weight, and the assumed bead pack modulus and the measured vertical resonant frequency, 27Hz, of the structure.

For quasi-static conditioning, an extra $\Delta M = 1$kg mass is added and then removed, three successive times, from the top of the structure. The resulting strain in the bead pack is estimated from this weight and the inferred modulus of the bead pack. $\epsilon = \Delta M g / A \rho c^2 = 2.35 \times 10^{-6}$ ($A$ being the area $\pi R^2$ with $R = 0.0355$m). Furthermore, because Hertzian theory has it that speeds should scale with (static load)$^{1/6}$, a final stretch after loading of $(1/6)(\Delta M / 44 kg) = 4 \times 10^{-3}$ is expected. As previously [11], we remark that an extrapolation to that final stretch leads to an expected time to full equilibrium greater than the age of the universe. See Table 1 in the main text for a summary and comparison with the glass bead pack [11].

*Single aluminum bead system*

Like the bead pack, the confined aluminum bead and the region of the plates near it are mechanically pumped in three different ways: impulsively, harmonically with a many-cycle sinusoidal force, and quasi-statically.

The impulsive conditioning was performed in two ways: i) dropping a 7g rubber ball from a height of 10cm, and ii) dropping a 0.72g wooden ball from a height of 21cm. Accelerometers on the upper slab, over the bead and over the support points, allow us to infer the corresponding transient force on the bead and thus the peak of the strain across the bead. For the rubber ball drop, $\epsilon_{peak} = 1360 \times 10^{-6}$, while for the wooden ball drop, $\epsilon_{peak} = 170 \times 10^{-6}$.

For harmonic conditioning, an electro-magnetic shaker was placed on the bottom slab and driven at 200Hz for three 2min intervals. Accelerometers again on the upper slab, over the bead and over the support points, allow us to infer the corresponding harmonic force on the bead and the harmonic strain amplitude across the bead: $\epsilon_{rms} = 44 \times 10^{-6}$.

Quasi-static conditioning consisted in alternately placing and removing a 65g weight on the upper plate over the bead. Based on Hertzian theory for the stiffness of the contacts, the corresponding quasi-static strain changes across the bead are $\Delta \epsilon = 110 \times 10^{-6}$. See Table 2 in the main text for a summary and comparison with the single glass bead system [12].

*Single steel bead system*

The confined steel bead is mechanically conditioned in two ways: impulsively, and quasi-statically.

Just as with the aluminum single bead system, the impulsive conditioning was performed in two ways, rubber ball drop and wooden ball, and quasi-static conditioning involved alternately adding and removing a 65g weight above the bead. The impulsive conditioning by the 7g rubber ball, dropped from 10cm, resulted in an estimated peak strain across the steel bead of $\epsilon_{peak} = 470 \times 10^{-6}$. The 0.72g wooden ball dropped from 21cm resulted in $\epsilon_{peak} = 190 \times 10^{-6}$. Quasi-static corresponded to a strain change of $\Delta \epsilon = 60 \times 10^{-6}$. See Table 2 in the main text for a summary.

**Appendix II: Aluminum bead pack under 215kPa load**

In addition to the observation of slow dynamics in an aluminum bead pack under a 109kPa static load, we also observe slow dynamics when the static load is increased to 215kPa (keeping everything else, e.g., the aluminum beads, the same). The slow dynamic recoveries are of similar magnitude but less clean than those shown in Figs. 1-3 of the main paper or those of the glass bead pack [11] at 215kPa. At 215kPa, the Al bead pack exhibits many slips (apparent as tiny steps in stretch followed by their own slow dynamic recoveries) minutes or more after the pump, and these contaminate the stretch versus time plots. It is also plausible to attribute the difference to extra plastic flow at the contact points. Figure S4 and S5 show the slow dynamic response to impulsive and quasi-static conditioning, respectively. The conditioning was the same as in the main paper.

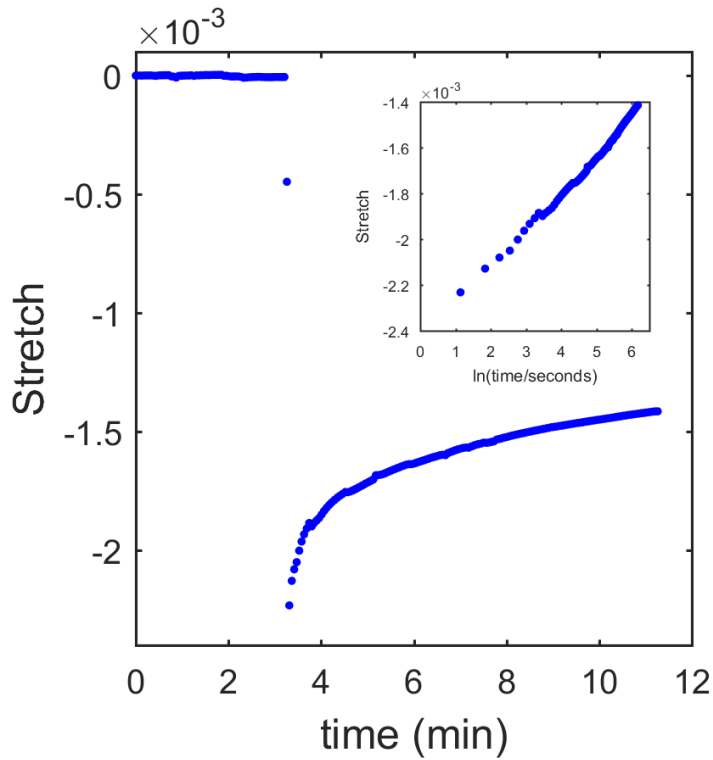

Figure S4. Impulsive conditioning on an aluminum bead pack under 215kPa static load. The bead pack was identical to the one discussed in Sec. II of the main paper except for the larger static load. The conditioning mechanism (rubber ball dropped from 20cm) is the same as Fig. 1. As opposed to the recovery in Fig. 1, many slips are observed here, possibly attributable to extra plastic flow at the contact points, as the static load is double that of Fig. 1.

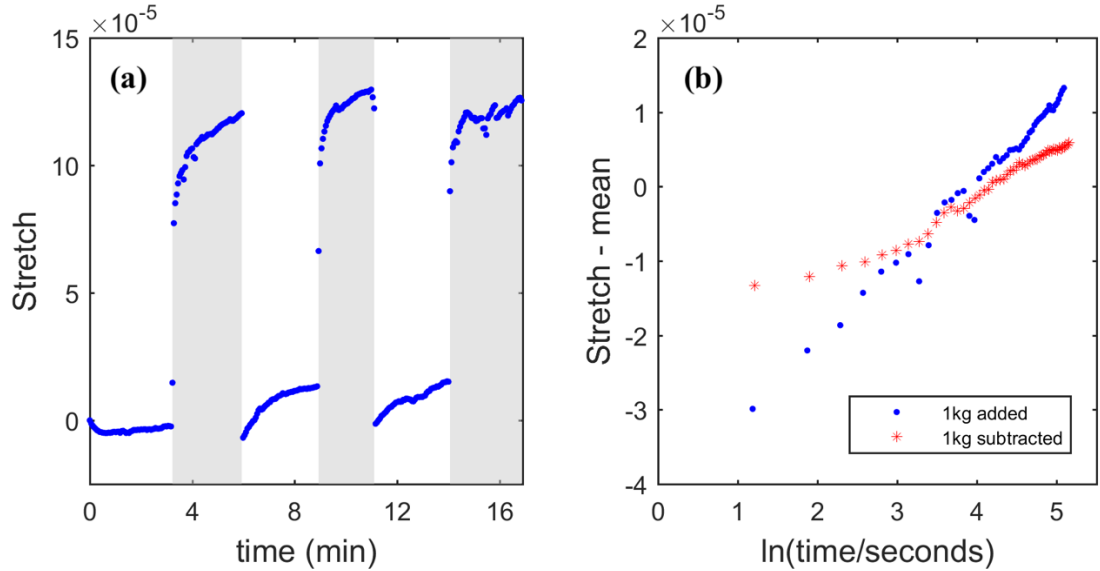

Figure S5. Quasi-static conditioning on an aluminum bead pack under 215kPa static load. The conditioning mechanism is the same as Fig. 2b (a 1kg weight was added and subtracted from the load). Shaded regions denote when the extra weight was added. As in the impulsive case, Fig. S4, slips are observed here.

**Appendix III: Impulsive conditioning with small wooden ball on single bead systems**

For both single bead systems (aluminum and steel), impulsive conditioning was performed in two ways: i) dropping a 7g rubber ball from a height of 10cm, and ii) dropping a 0.72g wooden ball from a height of 21cm. For the single glass bead system [12], impulsive conditioning involved dropping a 0.09g wooden ball from 21cm. The larger balls were necessary for good signal to noise because we lacked the vibration isolation used previously. Results for conditioning with the 7g rubber ball drop are presented in the main paper (Fig. 4), as those results had a better signal-to-noise than the 0.72g wooden ball. Here, we display the recoveries of the single aluminum bead system (Fig. S6) and the single steel bead system (Fig. S7) after impulse from the 0.72g wooden ball.

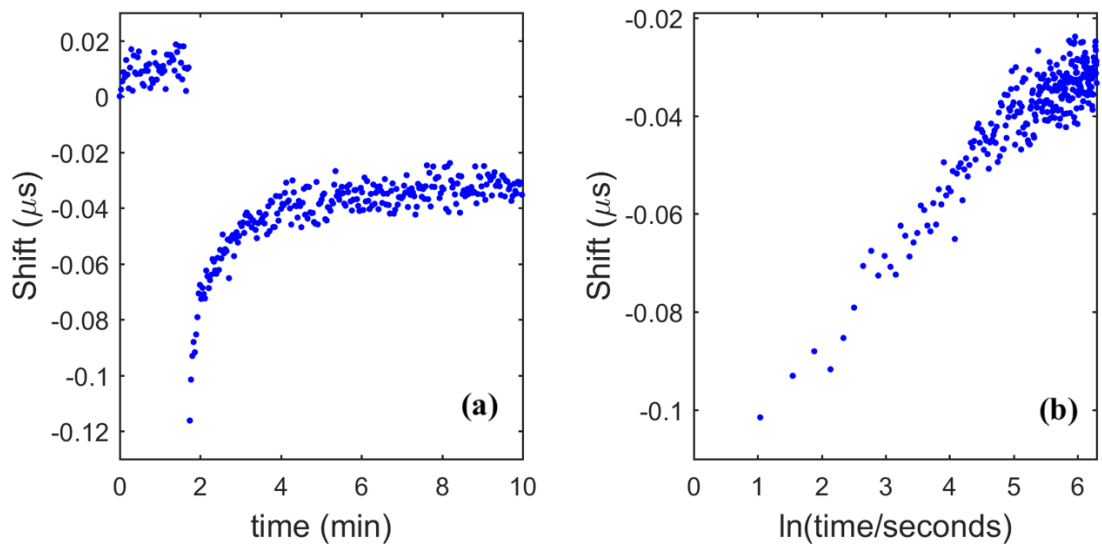

Figure S6. Slow dynamic response in the single aluminum bead system to impulsive conditioning from a 0.72g wooden ball drop.

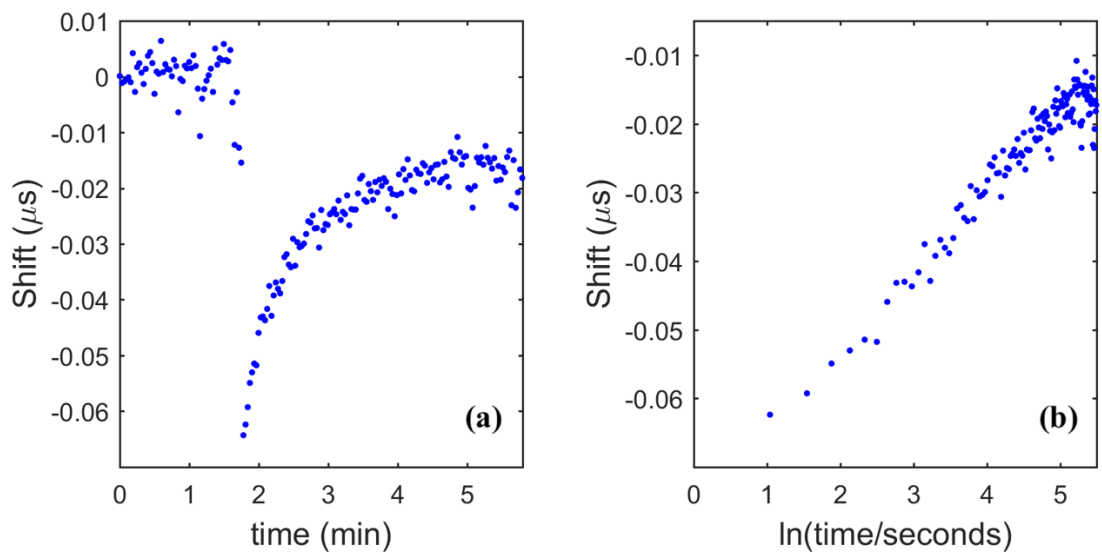

Figure S7. Slow dynamic response in the single steel bead system to impulsive conditioning from a 0.72g wooden ball drop.